\documentclass{PoS}

\title{The Pervasive Role of the Nuclear Symmetry Energy in the Structure
  and Evolution of Neutron Stars}

\ShortTitle{Pervasive Role of the Nuclear Symetry Energy}

\author{\speaker{M. Prakash} \\ 
  Department of  Physics and Astronomy, Ohio University \\ 
  Athens, Ohio 45701-2979, USA \\ 
E-mail: \email{prakash@harsha.phy.ohiou.edu}}

\author{{PALS}\thanks{MP thanks his ``PALS'' Paul Ellis (in memoriam),
James Lattimer, Sergey Postnikov, and Andrew Steiner who have helped
and contributed significantly to this talk.  Research support of the
US Department of Energy under grant DE-FG02-93ER40756 is gratefully
acknowledged. }}

\abstract{The multifaceted role of the density dependent nuclear
          symmetry energy in the nuclear astrophysics involving
          neutron stars is highlighted.  Efforts toward a model
          independent determination of the dense matter equation state
          through a deconstruction of the neutron star structure
          equation utilizing the masses and radii of several
          individual neutron stars are described. The need for
          observational data of {\em both} measurements for the same
          star is stressed.}

\FullConference{10th Symposium on Nuclei in the Cosmos\\
		 July 27 - August 1 2008\\
		 Mackinac Island, Michigan, USA}

\begin{document}

\section{Scope of This Talk}
In this talk, I will  
\begin{enumerate}
\item summarize some theoretical advances made in establishing correlations
  between observables of nuclei and neutron stars, 
\item indicate how laboratory experiments, e.g., at JLab and Rare
  Isotope Accelerators (RIA's), can help to unravel the composition
  and structure of a neutron star, 
\item point out how the observed masses ($M'$s) and radii ($R'$s) of
  several individual neutron stars can be used to construct the model
  independent equation of state (EOS) of dense matter (through an
  inversion process that I term as ``Deconstructing a Neutron Star''),
  and
\item indicate the key neutron star observations that are necessary to
  take leaps in our understanding of the nature of strong interactions
  at supra-nuclear densities.
\end{enumerate}

\section{The Pervasive Role of the Nuclear Symmetry Energy}
The nuclear (a)symmetry energy
\begin{equation}
E_{sym}(n,\delta) 
= \frac {1}{2n} \frac {\partial^2 \epsilon(n,x)}{\partial \delta^2}\,,
\qquad \delta = \frac {n_n - n_p}{n=n_p+n_n} = 1 -2x \,,
\end{equation}
represents the energy cost required to create an  isospin asymmetry
$\delta$ in nucleonic matter. Above, $n_n$ and $n_p$ are the neutron
and proton densities, $x$ is the proton fraction in matter, and
$\epsilon(n,\delta)$ is the energy density of isospin asymmetric
matter. Figure 1 from Ref.~\cite{Steiner05} shows the energy per
particle of bulk nuclear $(x=1/2)$ and neutron matter $(x=0)$ versus
the total baryon density $n$.
\begin{figure}[htbp]
\begin{center}
\includegraphics[width=2.75in]{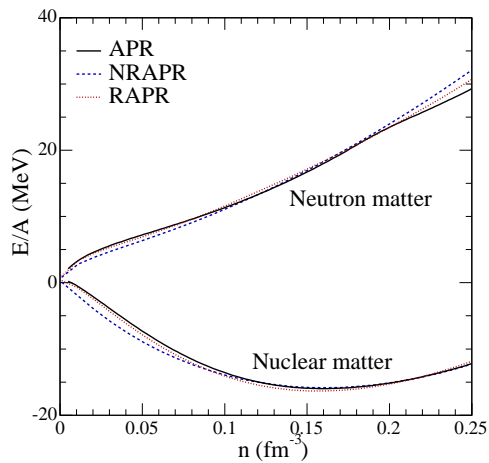}
\caption{
Results shown are for the microscopic potential model calculations of
Akmal and Pandharipande (APR)~\cite{AP,APR} and simpler
nonrelativistic and relativistic effective model fits~(NRAPR and
RAPR)~\cite{Steiner05}.}
\label{EOA}
\end{center}
\end{figure}

\begin{figure}[htbp]
\begin{center}
\includegraphics[width=6in]{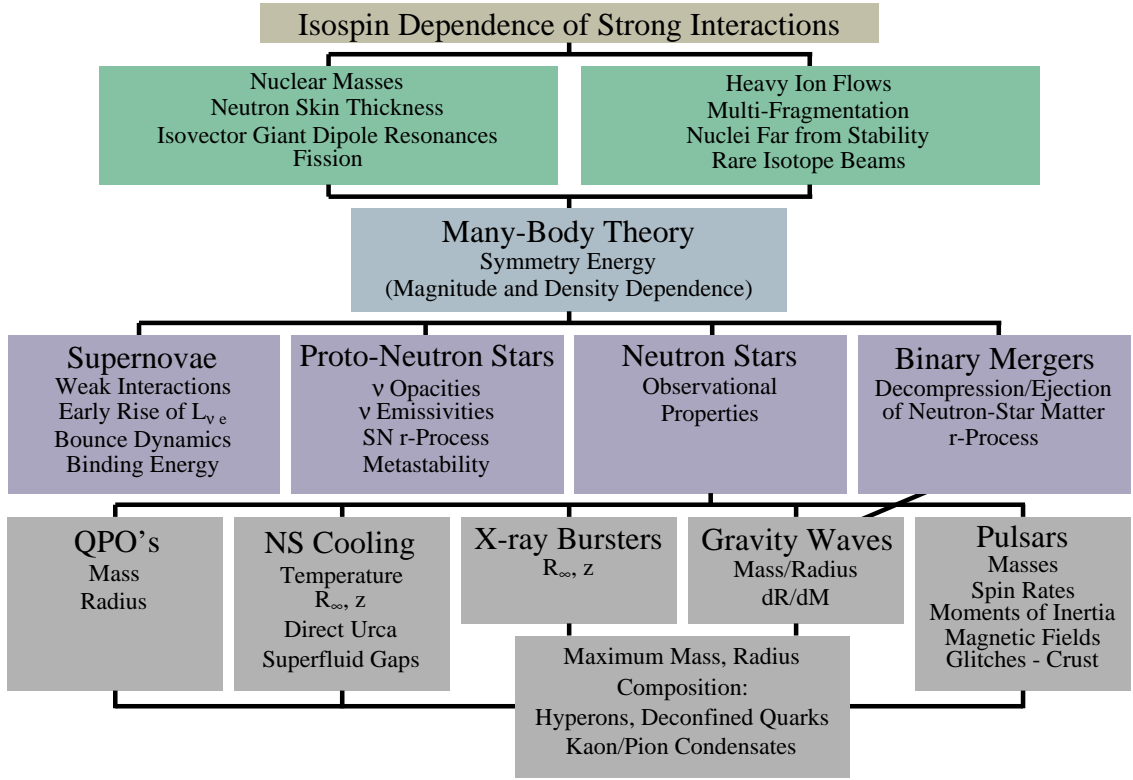}
\caption{
The pervasive role of the nuclear symmetry energy. Figure taken from
Ref.~\cite{Steiner05}.}
\label{corp}
\end{center}
\end{figure}

The physical properties of nuclei, such as their masses, neutron and
proton density distributions (including their mean radii), collective
excitations, fission properties, matter and momentum flows in high
energy heavy-ion collisions, etc. all depend on the isospin structure
of the strong interactions between nucleons (i.e., $nn$ and $pp$
interactions versus $np$ interactions).  The energy $\hat{\mu} = \mu_n
- \mu_p \cong 4 E_{sym}\delta$, where $\mu_n$ and $\mu_p$ are the
neutron and proton chemical potentials, respectively, is crucial in
determining reaction rates involving electrons and neutrinos, particle
abundances, etc., in astrophysical contexts such as supernova
dynamics, proto-neutron star evolution, the $r-$process, the long-term
cooling of neutron stars, and the structure of cold-catalyzed neutron
stars (i.e. their masses, radii and crustal extent), etc.  The
pervasive role of the isospin dependence of strong interactions in
nuclear processes in the laboratory and the cosmos is sketched in
Fig.~\ref{corp}.

\section{Connections to Neutron Star Structure}

\begin{figure}[htbp]
\begin{center}
\includegraphics[width=3.5in,angle=90]{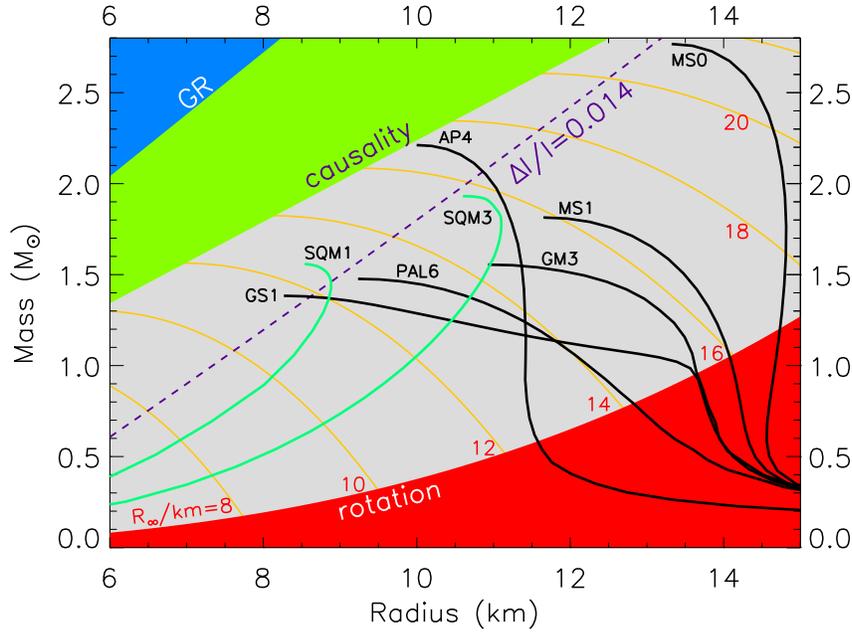}
\caption{
Mass versus radius curves from Ref.~\cite{LP04}.  EOS symbols are as
in Ref. ~\cite{LP01}.}
\label{mrcurves}
\end{center}
\end{figure}

The structure of neutron stars is determined by the energy and
pressure of charge-neutral beta-stable matter, the nucleonic
components of which can be written as
\begin{eqnarray}
E(n,x) &\cong & E(n,0.5) + E_{sym}(n)~(1-2x)^2 + \cdots \,, \nonumber \\
P(n,x) &\cong & n^2\left[E^\prime(n,0.5) + E^\prime_{sym}~(1-2x)^2 
\right] + \cdots\,,
\label{epnuc}
\end{eqnarray}
where the primes above refer to derivatives with respect to density.
Leptonic (electrons and muons) contributions must be added to
Eq.~(\ref{epnuc}) for the total energy and pressure. As demonstrated
in Ref.~\cite{LP01}, the radius (but, not necessarily the mass) of a
nucleonic neutron star is chiefly determined by the pressure in the
vicinity of the nuclear equilibrium density, $n_0 \simeq 0.16\pm
0.01~{\rm fm}^{-3}$, for which the contribution of the first term to
the pressure in Eq.~(\ref{epnuc}) is very small.  This feature of a
neutron star's structure highlights the importance of the density
dependence of the nuclear symmetry energy near $n_0$.  Figure
\ref{mrcurves} shows the mass-radius relationships for select EOS
models including the cases in which extreme softening due to the
presence of exotica such as hyperons, quarks or Bose condensates
occurs in the cores of stars.

\begin{figure}[htbp]
\begin{center}
\includegraphics[width=4in]{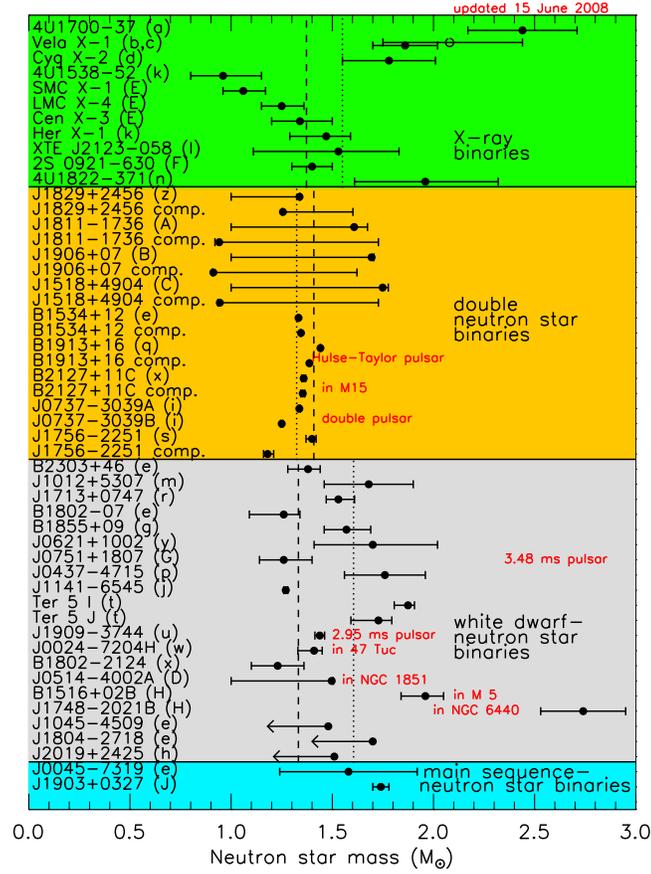}
\caption{
Observed neutron star masses. Figure courtesy of Lattimer. }
\label{masses}
\end{center}
\end{figure}

In Fig. \ref{masses}, the observed neutron star masses are shown. The
simple and error weighted means for neutron stars in x-ray binaries
are 1.55~${\rm M}_\odot$ and 1.37~${\rm M}_\odot$, respectively. The
corresponding numbers for double neutron star binaries are 1.32~${\rm
M}_\odot$ and 1.41~${\rm M}_\odot$, whereas for white dwarf - neutron
star binaries, the numbers are 1.60~${\rm M}_\odot$ and 1.33~${\rm
M}_\odot$.  It is unfortunate that in those cases (e.g., binary radio
pulsars) in which the masses are very accurately known, the
corresponding numbers for radii are unknown. It is hoped that with
improvements in timing analysis, the moment of inertia of the neutron
star in the double-pulsar system J0737-3039 will become
available~\cite{LS05}. As will be shown later, the masses and radii of
several individual neutron stars can uniquely pin down the EOS of
dense matter. For prospects of obtaining this much needed information,
see the article by Bob Rutledge in these proceedings. For prognosis of
EOS determination, see Ref.~\cite{LP07}.

\section{The Role of Isospin Interactions in Nuclei and Neutron Stars}

A few of the recently discovered empirical relationships that
underscore the role of isospin interactions in nuclei and neutron
stars are highlighted here.

\subsection*{The neutron skin thickness in nuclei and the sub-nuclear 
pressure of pure neutron matter} 

Typel and Brown \cite{Brown00,Typel01} have noted that model
calculations of the difference between neutron and proton rms radii
$\delta R ={\langle r_n^2\rangle}^{1/2} - {\langle
r_p^2\rangle}^{1/2}$ are linearly correlated with the pressure of pure
neutron matter at a density below $n_0$ characteristic of the mean
density in the nuclear surface (e.g., 0.1 fm$^{-3})$.  The density
dependence of the symmetry energy controls the so-called neutron skin
thickness $\delta R$ in a heavy, neutron-rich nucleus.  Explicitly,
$\delta R$ is proportional to a specific average of
$[1-E_{sym}(n_0)/E_{sym}(n)]$ in the nuclear surface, see
Refs.~\cite{Krivine84,Steiner05}.  

\begin{figure}[htbp]
\begin{center}
\includegraphics[width=2.75in]{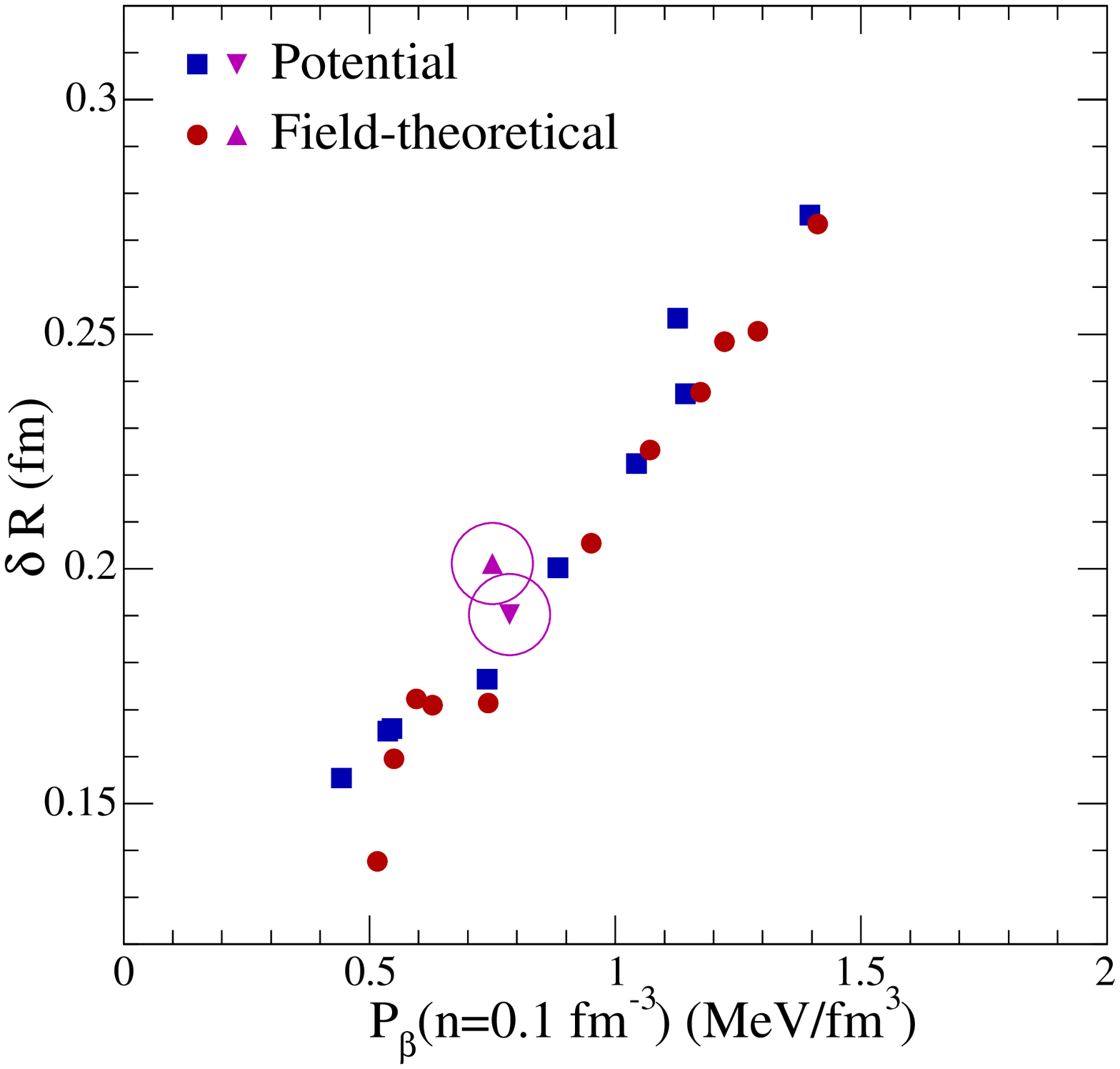}
\includegraphics[width=3in]{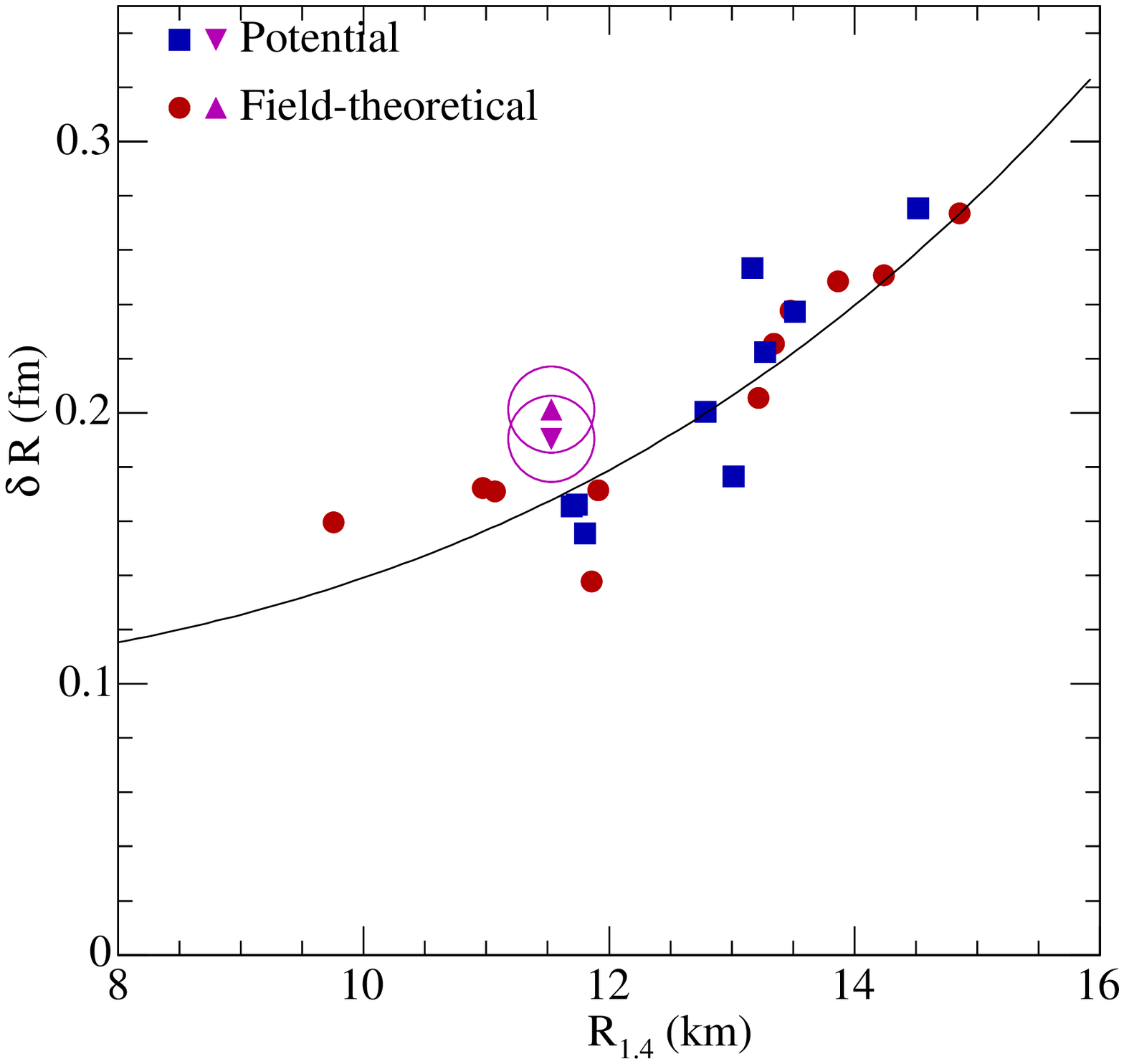}
\caption{
Left panel: The neutron skin thickness $\delta R$ of finite nuclei
versus the pressure of $\beta$-equilibrated neutron star matter at a
density of $0.1~{\rm fm}^{-3}$. Right panel: Calculated neutron skin
thicknesses $\delta R$ of nuclei versus the radii of $1.4{\rm M}_\odot$
stars. Trends for stars up to the maximum mass are similar.  The solid
lines represents a least square fit.  Trends for stars up to the
maximum mass are similar~\cite{Steiner05}. Figure taken from
~\cite{Steiner05}.}
\label{Typel}
\end{center}
\end{figure}

The left panel of Fig.~\ref{Typel} shows the Typel-Brown correlation
extended to neutron-star matter in Ref.~\cite{Steiner05} using both
potential and field-theoretical models.  The
proposal~\cite{Horowitz01b,Michaels00} at JLab to measure the neutron
rms radius of $^{208}$Pb through parity-violating electron scattering
experiments to 1\% accuracy (the rms charge radii of nuclei are known
to better accuracy) can help to provide a calibration pressure of
neutron-star matter at sub-nuclear densities.  The knowledge of
neutron rms radii for neutron-rich nuclei of varying masses will be of
great help in establishing the uncertain density dependence of isospin
interactions at sub-nuclear densities. The right panel of Fig. 5 shows
that the correlation between $\delta R$~ and~ $R_{1.4}$ is not very
sharp.  Accurate predictions of neutron star radii require knowledge
of the equation of state at supra-nuclear densities.

\subsection*{The neutron skin thickness in nuclei and the neutron star radius}

Horowitz and Piekarewicz \cite{Horowitz01} have pointed out that
models that yield smaller neutron skins in heavy nuclei tend to yield
smaller neutron star radii. These authors, along with others
\cite{Furnstahl02,Steiner05}, have stressed the need for accurate
measurements of the neutron skin thicknesses of nuclei. Although the
neutron star radii are determined at supra-nuclear densities, the calibration
provided at sub-nuclear densities cannot be ignored. Even with varying
stiffness at high density, several works have confirmed the trend
shown in the right panel of Fig.~\ref{Typel}. The Horowitz-Piekarewicz
correlation is thus very suggestive inasmuch as radii of stars up to
the maximum mass exhibit similar trends. It would be fruitful
to test this correlation from accurate radius measurements of neutron
stars whose corresponding masses are also well determined.

\subsection*{The neutron star radius $R$ and the pressure $P$ of 
neutron-star matter}

\begin{figure}[htbp]
\begin{center}
\includegraphics[width=3.5in]{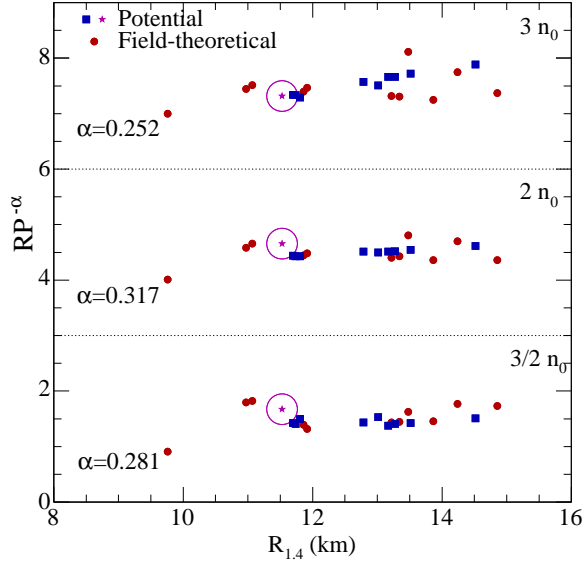}
\caption{The quantity $RP^{-\alpha}$ as a function of the stellar
radius $R$, for pressures $P$ determined at 3/2, 2 and 3 times
equilibrium nuclear matter density.  For each density, the best-fit
value for the exponent $\alpha$ is as indicated.  Circled stars
indicate the results obtained with the APR equation of state.  These
results are for $1.4{\rm M}_\odot$ stars; results for stars up to the
maximum mass star are similar. Figure taken from Ref.~\cite{Steiner05}.}
\label{latpra}
\end{center}
\end{figure}

Lattimer and Prakash \cite{LP01} found that the quantity $RP^{-1/4}$
is approximately constant, for a given neutron star mass, for a wide
variety of EOS's when the pressure $P$ of beta-equilibrated
neutron-star matter is evaluated at a density in the range $n_0$ to
$2n_0$, where $n_0$ denotes equilibrium nuclear matter density. Since
the pressure of nearly pure neutron matter (a good approximation to
neutron star matter) near $n_0$ is approximately given by
$n^2E_{sym}^\prime$ (see Eq.~(\ref{epnuc})), the density dependence of
the symmetry energy just above $n_0$ will be a critical factor in
determining the neutron star radius.  Figure \ref{latpra} shows
results for $1.4M_\odot$ stars. Similar trends persist for stars up to
the maximum mass star as shown in Ref.~\cite{LP01}.  This correlation,
coupled with accurate measurements of neutron star radii, can delimit
the very uncertain (up to a factor of 5!) pressure of neutron-star
matter (see Ref.~\cite{LP01}) at supra-nuclear densities.

\section{Deconstructing a Neutron Star}

This section contains a report of on-going work of the speaker with
Sergey Postnikov and James Lattimer~\cite{PPL}. The question being
posed here is ``Can observations of the masses and radii of several
(say 5 to 7) individual stars uniquely determine the dense matter
EOS?''  In short, the answer is a resounding YES! This assertion
follows from the work of Lindblom~\cite{Lindblom} who exploited the
one-to-one correspondence between an EOS and the $M-R$ curve generated
through the use of the Tolman-Oppenheimer-Volkov
(TOV)~\cite{Tolman,OV} equations of stellar structure. Given the
diverse theoretical predictions, it is of great interest to enquire
whether or not observations of masses and radii of several neutron
stars can be used to uniquely pin down the poorly known high density
EOS. The extent to which inherent errors in measurements affect the
extraction of the EOS (not addressed in Lindblom's work) will become
an important issue.

Recently launched campaigns~\cite{Rutledge08} to accurately measure
masses and radii of a large number of neutron stars give impetus to
investigate this inversion procedure in more depth than originally
undertaken by Lindblom.  The extension of such an inversion procedure
to include other observables such as the surface redshifts, binding
energies, moments of inertia, etc. has been undertaken by Postnikov,
Prakash and Lattimer incorporating the inherent errors involved in
observations.

We begin by recasting the TOV equations of stellar structure using
the variable $h$ defined through $dh = dp/(p+\rho(p))$, where $p$ is the
pressure and $\rho(p)$ is the mass-energy density. Adopting the units $G=1$ and
$c=1$, the result is
\begin{equation}
\frac {dr^2}{dh} = -2 r^2 \frac {r-2m}{m+4\pi r^3 p}\,, \qquad 
\frac {dm}{dh} = -4\pi r^3 \rho \frac {r-2m}{m+4\pi r^3 p}\,,
\label{ntov}
\end{equation}
where $p(h)$ and $\rho(h)$, which serve as input, contain the EOS.
The advantages of this reformulation (somewhat different than that used
by Lindblom) are that the enclosed mass $m$ and radius $r$ are now
dependent (on $h$ and thus the EOS) variables, and $h$ is finite both
at the center and surface of the star. Furthermore, the quantities
$r^2(h)$ and $m(h)$ admit tractable Taylon expansions about $h_c$ that
feature the finite quatities $p_c$, $\rho_c$ and $(d\rho/dh)_c$,
where the subscript $c$ denotes the star's center.

The procedure is to begin with a known EOS up to a certain
density, take small increments in mass and radius, and adopt an
iterative scheme based on Eq.~(\ref{ntov}) to reach the new
known mass and radius. In the absence of observational numbers for individual
stars, a test of the above scheme can be performed by assuming
that the masses and radii are those that result  from a model EOS.

\begin{figure}[htbp]
\begin{center}
\includegraphics[width=200pt,angle=0]{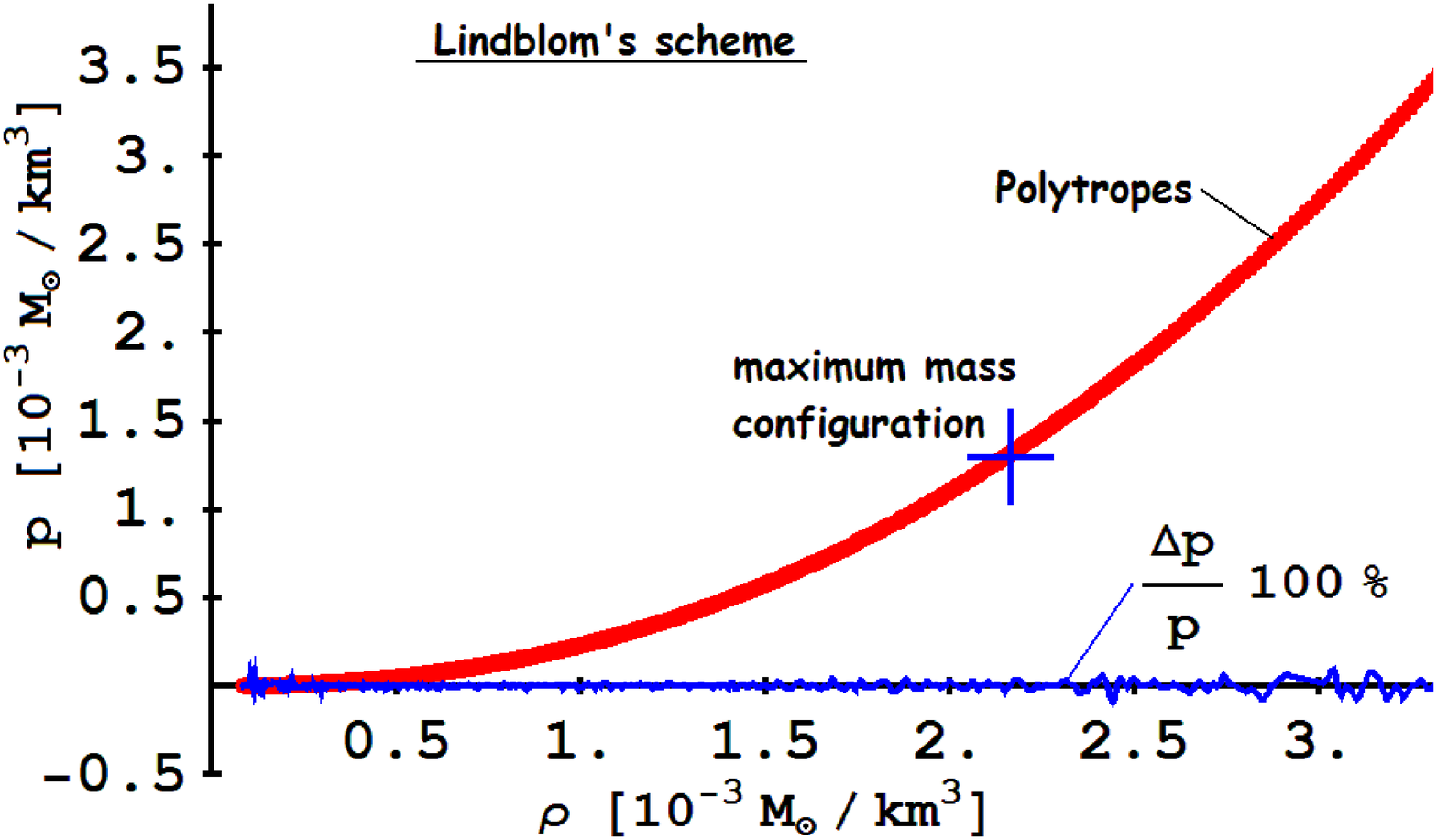}
\includegraphics[width=200pt,angle=0]{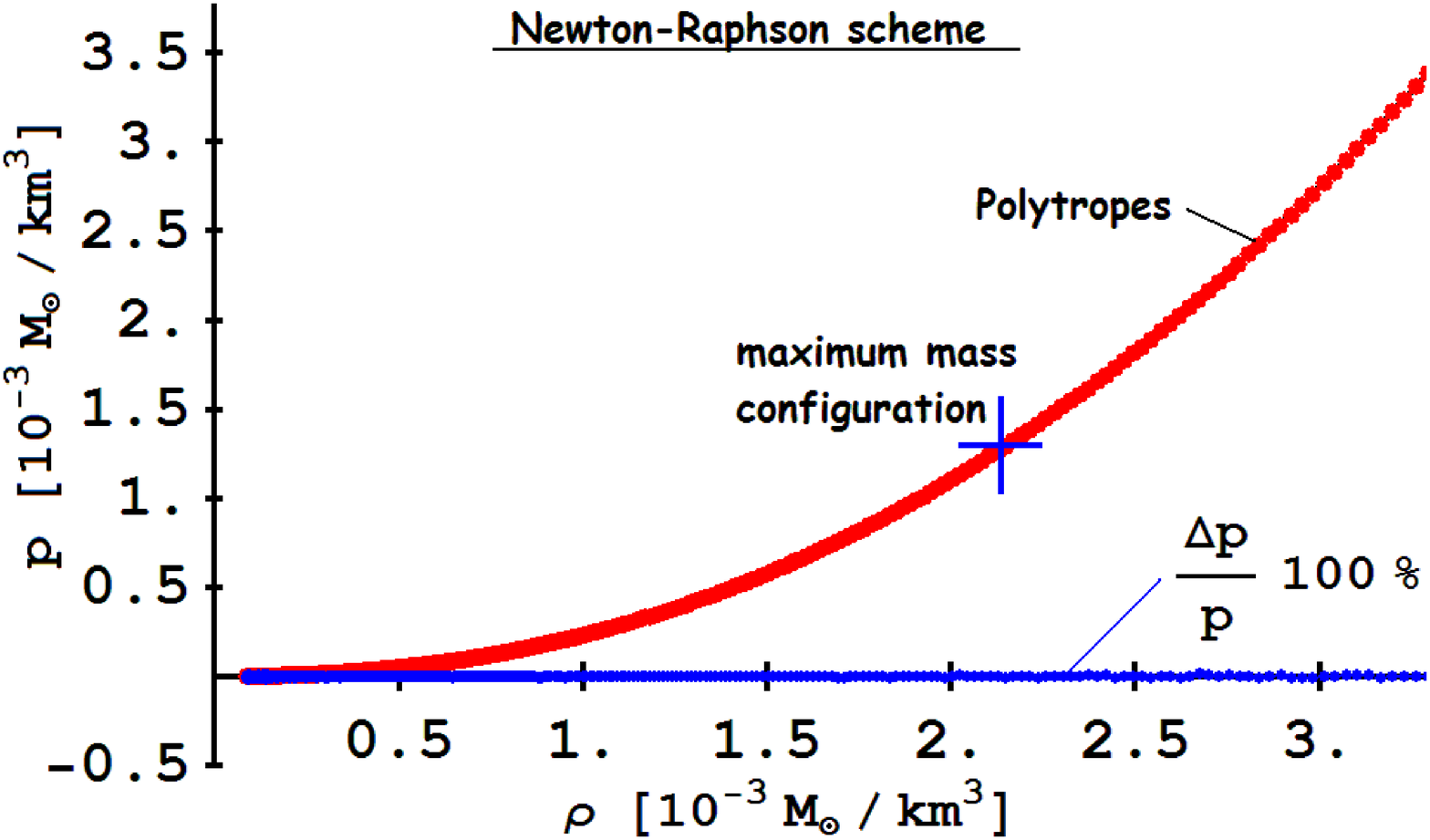}
\caption{Deconstructing a neutron star with a two-power polytropic EOS.}
\label{D1}
\end{center}
\end{figure}

We have tested our scheme using an EOS consisting of two
polytropes whose indices are appropriately chosen to mimic a realistic
EOS and obtained results that reproduce the exact EOS to hundredths of
percent accuracy.  The left panel of Fig. \ref{D1} shows our results
for this test case.  Methods to yield even better accuracy using this
modified Lindblom procedure are being devised. 

An alternative procedure is to solve the reformulated TOV equations
from the center to the surface with the assumed form of the EOS using
a Newton-Raphson scheme to obtain the known mass and radius.  The
results from this approach is shown in the right panel of
Fig. \ref{D1}. This method also yields results to hundredths of
percent accuracy. We now have two methods that yield similarly
accurate results.

\begin{figure}[htbp]
\begin{center}
\includegraphics[width=200pt,angle=0]{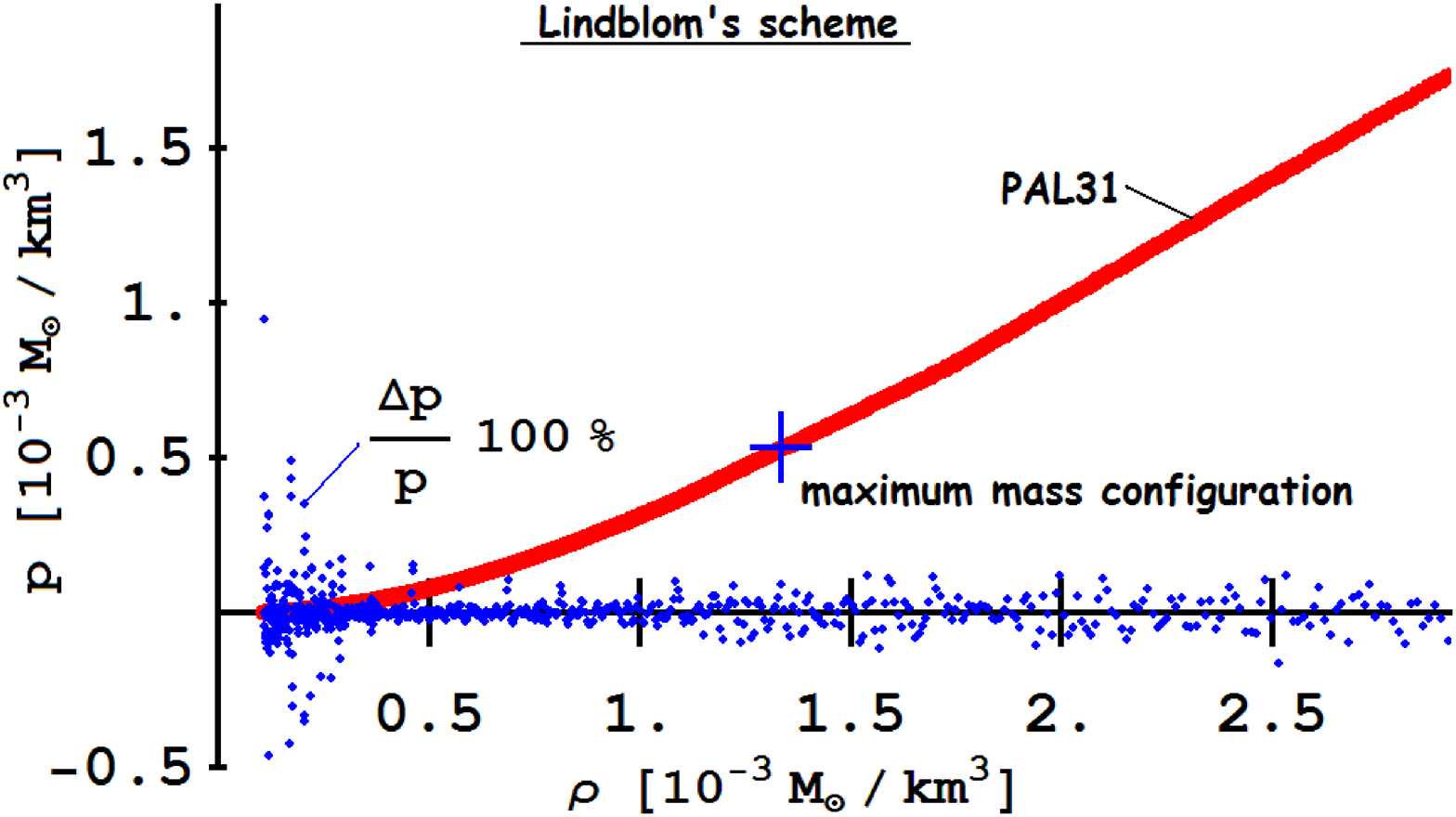}
\includegraphics[width=200pt,angle=0]{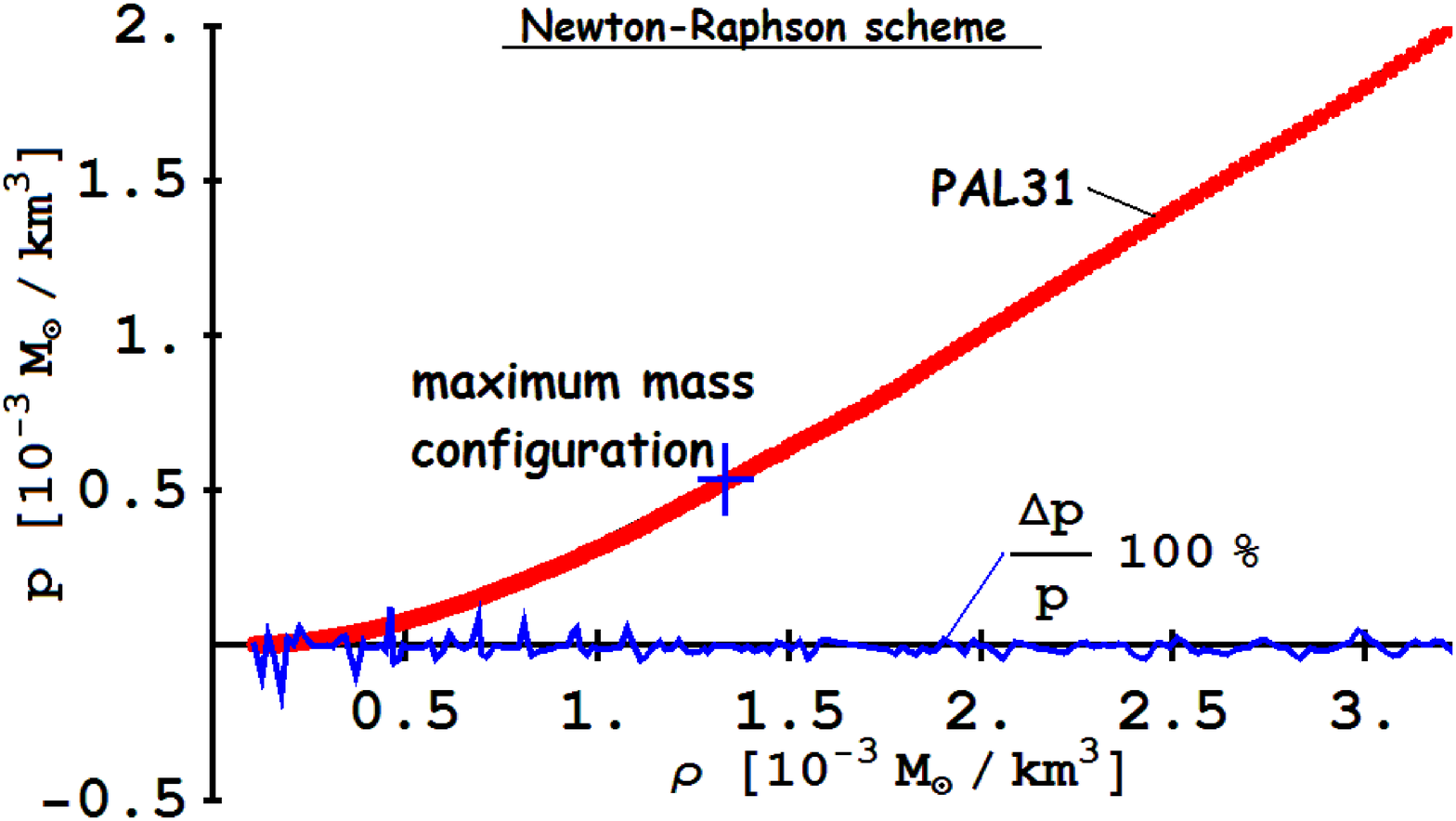}
\caption{Deconstructing a neutron star with a physically motivated
nucleonic EOS.}
\label{D2}
\end{center}
\end{figure}

We are now in a position to test our two schemes using physically
motivated model EOS's.  As examples of realistic EOS's, we will begin
with the EOS's of Ref.~\cite{PAL} . These EOS's allow for variations
in the uncertain high-density behavior of the nuclear symmetry energy
and mimic trends found in microscopic calculations of high-density
matter. The EOS labeled PAL31 is characterized by the nuclear matter
compression modulus of $K=240$ MeV and a symmetry energy function for
potential interactions: $F(u)=u$, where $u=n/n_0$ and $n_0=0.16~{\rm
fm}^{-3}$. Including contributions from kinetic energy, the symmetry
energy at $n_0$ is 30 MeV.  The results in Fig. \ref{D2} show the
extent to which the EOS is reconstructed for the case in which the
masses and radii are assumed to be those that result from the EOS of
PAL31~\cite{PAL}.  The two iterative schemes described above yield
results of similar satisfactory accuracy.

Work is in progress to determine the extent to which 
``Deconstruction'' works accurately 
for a variety of EOS's including those that permit
extreme softening at high density. Several exercises can
be envisaged: (i) To what extent can one determine if only one mass
and radius are very well determined? (ii) What if several masses and
radii are known, albeit with unsatisfactory error bars? (iii) To what
extent can we determine the EOS given the inherent error bars in
measurements?. Answers to these questions and many others will be
published soon~\cite{PPL}.  In addition to its inherent worth, this
investigation will aid astronomers to better plan and secure
observations in the future.

\section{Conclusions}

It is clear that 

\begin{enumerate}

\item the masses and radii of several (say 5 to 7) of individual neutron
  stars can pin down (through deconstruction) the equation of state of
  neutron-star matter and shed light on the density dependence of the
  symmetry energy, i.e., the poorly known isospin dependence of strong
  interactions, and 

\item precise laboratory experiments, particularly those involving
  neutron-rich nuclei, are sorely needed to pin down the near-nuclear
  aspects of the symmetry energy which determines the masses, neutron
  skin thicknesses, collective excitations, etc., of nuclei. 

\end{enumerate}

It is also clear that continuing mutual interactions of astronomers,
laboratory experimenters and theorists in the fields of astrophysics
and strong interaction physics are necessary to fulfill the objective
of understanding the physics of compact objects in which the ultimate
energy density of observable cold-baryonic matter is realized~\cite{LP05a}.


\begin{thebibliography}{99}

\bibitem{Steiner05}A. W. Steiner, M. Prakash, J. M. Lattimer and
  P. J. Ellis, Phys. Rep. {\bf 411} (2005) 325.

\bibitem{AP}
A. Akmal and V. R. Pandharipande, Phys. Rev. C {\bf
  58} (1997) 2261. 

\bibitem{APR}
A. Akmal, V. R. Pandharipande and D. G. Ravenhall, Phys. Rev. C {\bf
  65} (1998) 1804. 

\bibitem {LP01}J. M. Lattimer and M. Prakash, Astrophys. Jl. {\bf 550}
  (2001) 426. 

\bibitem {LP04}J. M. Lattimer and M. Prakash, Science {\bf 304} (2004) 536.

\bibitem{LS05}J. M. Lattimer and B. F. Schutz, Astrophys. Jl. {\bf
  629} (2005) 979.


\bibitem {LP07}J. M. Lattimer and M. Prakash, Phys. Rep. {\bf 442}
  (2007) 109.

\bibitem{Brown00}B. A. Brown, Phys. Rev. Lett. {\bf 85} (2000) 5296.

\bibitem{Typel01}S. Typel and B. A. Brown, Phys. Rev. C {\bf 64}
  (2001) 027302.
 
\bibitem{Krivine84}H. Krivine, J. de Phys. C {\bf 6} (Supp.) (1984)
  153. 

\bibitem{Horowitz01b}C. J.Horowitz, S. J. Pollock, P. A. Souder,
  R. Michaels, Phys. Rev. C {\bf 63} (2001) 025501.  

\bibitem{Michaels00}R. Michaels, P. A. Souder, G. M. Urciuoli,
  Proposal PR-00-003, Jefferson Laboratory, 2000.

\bibitem{Horowitz01}C. J. Horowitz and J. Piekarewica,
  Phys. Rev. Lett. {\bf 86} (2001) 5647.

\bibitem{Furnstahl02}R. J. Furnstahl, Nucl. Phys. {\bf A 706} (2002) 85.

\bibitem{PPL}S. Postnikov, M. Prakash and J. M. Lattimer, In preparation.

\bibitem{Lindblom}L.~Lindblom, Astrophys. Jl. {\bf 398} (1992) 569.

\bibitem{Tolman}R. C. Tolman, Proc. Nat. Acad. Sci., U. S. A., {\bf
  20} (1934) 3.

\bibitem{OV}J. R. Oppenheimer and G. M. Volkoff, Phys. Rev. {\bf 55}
  (1939) 374.

\bibitem{Rutledge08}See the contribution from Rutledge in these
  proceedings.

\bibitem{PAL}M. Prakash. T. L. Ainsworth and J. M. Lattimer, 
Phys. Rev. Lett. {\bf 61} (1988) 2518.

\bibitem{LP05a}J. M. Lattimer and M. Prakash, Phys. Rev. Lett. {\bf
  94} (2005) 111101. 

\end{thebibliography}
\end{document}